# Quantum Plasmonics

M. S. Tame[1*], K. R. McEnery[1,2], Ş. K. Özdemir[3], J. Lee[4], S. A. Maier[1*] & M. S. Kim[2]

[1] *Experimental Solid State Group, Department of Physics, Imperial College London, Prince Consort Road, SW7 2BW, United Kingdom*

[2] *Quantum Optics and Laser Science Group, Department of Physics, Imperial College London, Prince Consort Road, SW7 2BW, United Kingdom*

[3] *Department of Electrical and Systems Engineering, Washington University, St. Louis, MO63130, USA*

[4] *Department of Physics, Hanyang University, Seoul 133-791, Korea*

**Quantum plasmonics is a rapidly growing field of research that involves the study of the quantum properties of light and its interaction with matter at the nanoscale. Here, surface plasmons - electromagnetic excitations coupled to electron charge density waves on metal-dielectric interfaces or localized on metallic nanostructures - enable the confinement of light to scales far below that of conventional optics. In this article we review recent progress in the experimental and theoretical investigation of the quantum properties of surface plasmons, their role in controlling light-matter interactions at the quantum level and potential applications. Quantum plasmonics opens up a new frontier in the study of the fundamental physics of surface plasmons and the realization of quantum-controlled devices, including single-photon sources, transistors and ultra-compact circuitry at the nanoscale.**

Plasmonics provides a unique setting for the manipulation of light via the confinement of the electromagnetic field to regions well below the diffraction limit[1,2]. This has opened up a wide range of applications based on extreme light concentration[3], including nanophotonic lasers and amplifiers[4,5], optical metamaterials[6], biochemical sensing[7] and antennas transmitting and receiving light signals at the nanoscale[8]. These applications and their rapid development have been made possible by the large array of experimental tools that have become available in recent years for nanoscale fabrication and theory tools in the form of powerful electromagnetic simulation methods. At the same time and completely parallel to this remarkable progress, there has been a growing excitement about the prospects for exploring quantum properties of surface plasmons and building plasmonic devices that operate faithfully at the quantum level[9]. The hybrid nature of surface plasmon polaritons (SPPs) as 'quasi-particles' makes them intriguing from a fundamental point-of-view, with many of their quantum properties still largely unknown. In addition, their potential for providing strong coupling of light to emitter systems, such as quantum dots[10,11] and nitrogen-vacancy (NV) centres[12], via highly confined fields offers new opportunities for the quantum control of light, enabling devices such as efficient single-photon sources[13,14,15,16] and transistors[17,18,19] to be realized. While surface plasmons are well known to suffer from large losses, there are also attractive prospects for building devices that can exploit this lossy nature for controlling dissipative quantum dynamics[20]. This new field of research combining modern plasmonics with quantum optics has become known as 'quantum plasmonics'.

In this review, we describe the wide range of research activities being pursued in the field of quantum plasmonics. We begin with a short description of SPPs and their quantization. Then, we discuss one of the major strengths of plasmonic systems: the ability to provide highly confined electromagnetic fields. We describe how this enables the enhancement of light-matter interactions and the progress that has been made so far in demonstrating a variety of schemes that take advantage of it in the quantum regime. We also review key experiments that have probed fundamental quantum properties of surface plasmons and their potential for building compact nanophotonic circuitry. We conclude by providing an outlook on some of the important challenges that remain to be addressed and new directions for the field.

**Quantization**

One of the most fundamental aspects in quantum plasmonics is the description of surface plasmons using quantum mechanics. This is what sets it apart from all other areas of modern plasmonics. Much of the work laying the foundations for quantization was carried out in the 1950s by Bohm and Pines, with work by Pines providing the very first model for quantizing plasma waves in metals[21]. Here, electrons in the conduction band were considered to be free electrons in an electron gas and the long-range correlations in their positions treated in terms of collective oscillations of the system as a whole. The quantized form of these collective matter oscillations – plasmons – were found to be bosons, with both wave-like and particle-like behaviour, as expected for quantum excitations. The 'polariton' – a joint state of light and matter – was introduced by Hopfield[22], who provided a quantum model for the polarization field describing the response of matter to light. Depending on the type of matter, Hopfield called the field a 'phonon-polariton', 'plasmon-polariton' and so on, with the quanta as bosons. The concept of a surface plasma wave (SPW) was proposed soon after by Ritchie[23]. Several years later, Elson and Ritchie[24], and others used Hopfield's approach to provide the first quantized description of SPWs as 'surface plasmon polaritons', whose coupled light-matter features are described in Figure 1. Hydrodynamic effects were also included in the quantization[25]. Despite its great success, Hopfield's approach did not consider loss, which is caused by the scattering of electrons with background ions, phonons and themselves in the conduction band[8,26] (ohmic loss) and at high frequencies by interband transitions[26]. A new 'microscopic' quantization method was introduced by Huttner and Barnett[27], extending Hopfield's approach to polaritons in dispersive and lossy media, including waveguides. Most recently a 'macroscopic' approach has been developed using Green's functions[28]. Localized surface plasma oscillations at nanoparticles have also been quantized[29,30,31], the quanta of which are called localized surface plasmons (LSPs). In Box 1, we outline a basic approach to quantization for the waveguide[32,33] (SPP) and localised[30,31] (LSP) setting.

**Optical confinement**

The ability of SPPs to confine and guide their coupled light field within regions far below the diffraction limit is one of their major strengths first highlighted by Takahara *et al.*[1] Here, a nanowire with negative dielectric function, $\varepsilon$, was considered, where it was found that fundamental limits imposed on the field confinement for standard optical materials with positive $\varepsilon$ were no longer valid. Subsequent work[36,37] showed the underlying difference between subwavelength confinement, which standard optical materials can also achieve (using large positive $\varepsilon$), and subdiffraction confinement, which is a unique feature of light guided by SPPs and localised by LSPs using materials with negative $\varepsilon$, such as metals[26], superconductors[26] and graphene[26,38]. The principles of these two key concepts are described in Box 2.

By confining light using SPPs or LSPs, one is able to significantly alter the photonic density of states. Thus, the dynamics of light-matter interactions can be significantly modified and enhanced. Several groups initiated investigations into the emission of light from isolated matter systems placed close to metal. Most notably, Hecker *et al.* observed a 3-fold enhancement in the luminescence from a single quantum well and found that it was due to the generation of SPWs[39]. Neogi *et al.* used time-resolved photoluminescence measurements to demonstrate the enhancement of spontaneous emission due to the coupling of a quantum well to SPWs[40]. The enhancement was quantified using the Purcell factor[41] – the ratio of the spontaneous emission rate to that in free space – with values of up to 92 observed. These experimental works and related quantum optical models[42] provided a stimulating backdrop for researchers as they started to explore plasmonic systems using quantum optics techniques.

## Quantum Properties of SPPs

**Survival of entanglement**

The first experimental observation of quantum optical effects in a plasmonic structure was reported by Altewischer *et al.*[43] In this pioneering work (Figure 2a) it was shown that when one or both photons from a polarization-entangled pair are converted into SPPs on gold films perforated by subwavelength holes (gold grating), then back into photons, their entanglement survives. Although many incident photons were lost due to losses in the metal, the photons that survived and reached the detectors were found to be highly entangled. Since this experiment and its quantum description[44], many further experiments were reported, suggesting that entanglement in other degrees of freedom could also be transferred into a plasmonic structure, maintained and released back out. An experiment by Fasel *et al.* demonstrated that energy-time entanglement in a photon pair can be preserved during the photon-SPP-photon conversion using a gold grating and a long-range SPP (LRSPP) waveguide[45]. The preparation of energy-time entangled SPPs in two separate LRSPP waveguides was later reported[46], where coherence lengths shorter than the plasmon propagation distance were used, ensuring complete conversion of photons into SPPs. A single SPP in a coherent superposition of existing at two different times, with a delay much larger than the SPP lifetime, was also realized[46]. The work on entanglement was extended by Ren *et al.* to spatial modes[47], showing that entanglement of orbital angular momentum could also survive conversion into and out of SPPs. Guo *et al.* demonstrated the preservation of two-photon quantum coherence by sending two photons onto a gold grating, and finding that interference fringes at a Mach-Zehnder interferometer showed the distinct two-photon de Broglie wavelength before and after the plasmonic structure[48]. Huck *et al.* demonstrated the robustness of continuous-variable states during the photon-SPP-photon conversion process[49], hinting at the possibility of controlling continuous-variable quantum states using plasmonics.

These initial experiments confirmed that photonic entanglement and quantum information could be encoded into the collective motion of a many-body electronic system, and that the macroscopic nature of an SPP (involving $\sim 10^6$ electrons) does not destroy quantum behaviour. This was surprising as it was anticipated that the collisions in this massive collection of charges would inevitably lead to decoherence and the loss of quantum information.

**Decoherence and loss**

Although quantum properties of the light field were found to survive the photon-SPP-photon conversion process to a high degree, in several experiments[43,45,46] the visibilities of the interference fringes (first-order coherence) were found to decrease due to ohmic loss and surface scattering. In their study of squeezed states, Huck *et al.* successfully modelled the decoherence leading to the degradation of squeezing in the photon-SPP-photon conversion process as a beam splitter interaction[33,49]. Di Martino *et al.* went further by exciting single SPPs in metallic stripe waveguides of different lengths[50] (Figure 2b). They found that the losses incurred during propagation are consistent with an uncorrelated Markovian linear loss model. Fujii *et al.,* on the other hand, have revealed that in addition to linear loss, in quantum plasmonic systems there are chromatic and dispersion effects which may induce temporal and spectral mode distortion[51].

**Wave-particle duality**

One of the fundamental features of quantum mechanics is that a single quantum excitation exhibits both wave-like and particle-like behaviour. Kolesov *et al.* demonstrated wave-particle duality for SPPs[12] (Figure 2c). Here, single SPPs on a silver nanowire were generated by driving NV centres with an external field. The SPPs were found to self-interfere, clearly showing the wave-like behaviour. They then showed the particle-like behaviour via the measurement of the second-order quantum coherence function.

**Quantum size effect**

Depending on the size of a metal nanostructure, microscopic quantum effects can be significant in the description of the electrodynamics. The continuous electronic conduction band, valid at macroscopic scales, breaks up into discrete states when the dimensions are small enough, making the Drude model for the dielectric function no longer valid[52,53,55,56]. Many experiments have optically probed this quantum size effect[52,56,57], which manifests itself as a shift and broadening of the plasmon resonance, in addition to the appearance of a fine structure, corresponding to transitions between the discrete energy levels[52,58,59,60]. Here, electron-energy-loss spectroscopy with a scanning transmission electron microscope has been used[52,57]. Scholl *et al.* have found that as the diameter of a nanoparticle approaches a critical size, the plasmon resonance undergoes a blue shift with linewidth broadening, which is drastically different to the predictions of classical electromagnetism[62,63]. While analytical models are still used[53,54,55,61,62], recent work has employed numerical density functional theory (DFT) to model the many-body electron system, obtaining quantum corrected dielectric functions for predicting experimental observations. DFT accounts for the spill-out of electrons outside a nanoparticle and the gradual change of the dielectric properties at the surface. Using DFT, Prodan *et al.* have shown that the electron spill out in nanoshells can introduce new modes and a broadening of the plasmon resonances[64], in addition to strong changes in the plasmon line shapes due to the interplay between plasmons and single-electron excitations[65]. Zuloaga *et al.* have also used DFT to investigate quantum plasmonic behaviour in nanorods[66] and dimers[71]. Townsend and Bryant have found that in small nanospheres there can be two types of collective oscillations, quantum core plasmons in the centre and classical surface plasmons throughout[67]. Quantum size effects in thin films[68,69] and graphene[70] have also been studied. These works have shown that quantum size effects need to be taken into account when designing ultracompact nanophotonic devices based on plasmonics.

**Quantum tunneling**

When metallic nanostructures are placed close to each other quantum tunneling can occur. Zuloaga *et al.* have shown that electron tunneling effects can play an important role in the optical resonances between two nanoparticles with separation distances $d < 1$nm[71]. Moreover, for distances $d < 0.5$nm the dimer enters a conductive regime, where a charge transfer plasmon mode appears involving electrons flowing back and forth between the particles. Mao *et al.* have investigated quantum tunneling between two silver plates, showing that it is responsible for a reduction in surface-enhanced Raman scattering[72]. Savage *et al.* have experimentally revealed the quantum regime of tunneling plasmonics in subnanometer plasmonic cavities formed by two nanostructures[73] (Figure 2d). They found that as the nanostructure separation decreases below a critical size, the plasmon interactions enter the quantum regime, manifested by a blue shift of the resonances, attributed to the screening of localized surface charges by quantum tunnelling and a consequent reduction in the plasmonic coupling. The results agree well with the predictions of the quantum corrected model of Esteban *et al.*[74] and recent experiments by Scholl *et al.*[75] Nonlinear effects in quantum tunneling have also been investigated[76]. In a recent study[77], Wu *et al.* considered Fowler-Nordheim tunneling, which occurs in the presence of an external high electric field. Here, electrons from the conduction band of one nanoparticle tunnel into the gap between the nanoparticles and are swept into the other nanoparticle. This process occurs when the barrier has a sloped energy-space profile. The strength and damping rate of plasmonic oscillations can be controlled by tuning the intensity of the incident light. Thus, the charge transfer can be modulated by an external source, which may be useful for developing novel quantum devices such as switches.

## Single emitters coupled to SPPs

The large size mismatch between light and single emitters ensures that their light-matter interaction is inherently weak. This is a problem as strong, coherent coupling between single photons and emitters is critical for developing future quantum technology[78]. There are several strategies to circumvent this problem. High quality cavities have been used to boost interaction times and encourage stronger coupling. However the use of cavities places a restriction on the bandwidth and the size of devices. An alternative strategy is to use an interface to bridge the size gap. Confining the light field to small effective volumes in this way enables stronger coupling with the emitter. Plasmonic modes can be squeezed into volumes far below the diffraction limit, and therefore provide an excellent interface between single photons and emitters[10].

**Weak and strong coupling**

Light-matter interactions can be split into two principal regimes, the weak-coupling and the strong-coupling regime (Box 3). The weak-coupling regime is associated with the Purcell enhancement of spontaneous emission. This effect has been found to be particularly strong when an emitter is placed next to a metallic surface or nanostructure[84,85,86], where the emitter couples to confined plasmonic modes[87]. Plasmonic modes are able to strongly enhance the fluorescence of emitters despite having low quality factors due to ohmic losses. This enhancement is due to two simultaneous processes[88]. First, the intense plasmonic field increases the *excitation* rate of the emitter. Second, the subwavelength confinement of the light field enhances the *decay* rate of the emitter into the plasmonic mode via the Purcell effect[41] (see Box 3). The fluorescent enhancement is tempered by the non-radiative excitation of lossy surface waves at the metal surface[88]. This process, known as fluorescence quenching, occurs close to the surface and therefore leads to an optimal distance for coupling the emitter into a plasmonic mode. The high quality factors, Q, or long interaction times associated with traditional cavities limits the speed at which photons can be emitted once collected into the cavity. Plasmonics does not suffer from this problem and thus promises single-photon sources on chip at optical frequencies with high operation speed. This plasmon-induced Purcell enhancement can also be used to encourage quantum interference between the transitions of a multi-levelled emitter, leading to an enhancement in phenomena such as electromagnetic-induced transparency, coherent population trapping and lasing without inversion[89,90]. Additional effects predicted also include polarized resonance fluorescence[91], quantum beats of trapped populations[92] and narrowing the linewidth of spontaneous emission[93].

Recently, a 2.5-fold enhancement in the emission of a single quantum dot into an SPP mode of a silver nanowire was demonstrated by Akimov *et al.*[11] (Figure 3a). Moreover, they observed that the light scattered from the end of the nanowire was anti-bunched (Figure 3b), confirming that the SPP mode could collect and radiate single photons from the quantum dots. Subsequent experiments have shown Purcell enhancements of single emitters coupled to SPP[12,94,95,96,97,98] and LSP[88,99] modes. Further efforts have also been made to exploit more advanced designs to improve collection and control. One example is hybrid SPPs[100,101], where a waveguide gap is used to achieve Purcell factors as high as 60. The growing use of nanoantenna to control the emission direction of the collected light[102,103,104,105] is another example. These efforts point towards the exciting prospect of single-photon antennas[106] that can efficiently absorb light from emitters and subsequently emit the photons in a well-controlled manner.

The second principal regime is the strong-coupling regime. Here, the interaction between light and matter can be described by the coupling, $g \propto \sqrt{\frac{1}{V_{eff}}}$. While confined plasmonic modes couple very strongly to matter, unfortunately because of large ohmic losses it is not easy to enter the strong-coupling regime in plasmonic systems, where light-matter interactions must be dealt with non-perturbatively. There is, however, a regime where the coupling strength is intermediate between the mode and the emitter dissipation. This is known as the bad-cavity limit in cavity quantum electrodynamics (CQED) and displays interesting physics, such as cavity-induced transparency[107]. A similar effect has been studied in coupled metal nanoparticle-emitter systems, where very large enhancements in response have been predicted[31,108,109].

In general, the strong-coupling regime is characterized by the reversible exchange of energy between the light field and the emitter – Rabi oscillations. These oscillations manifest themselves in an energy splitting of the light-matter energy levels. There have been experimental observations of these splittings in the spectra of ensembles of molecules due to plasmonic interactions[110,111,112,113]. Experimental evidence for strong coupling between a single emitter and a plasmonic mode, however, is still elusive. Classical predictions have suggested strong coupling could be achieved between an emitter and a metallic dimer antenna[114]. There have also been theoretical examinations of the strong-coupling regime based on a fully quantum mechanical framework[30,115,116]. These works take into consideration higher order modes whose relevance cannot be ignored as the metal-emitter separation decreases past the point where the dipole approximation is valid. As a result, the intuitive CQED analogy[31] is replaced with macroscopic QED techniques better suited to more complex systems[35]. Trügler and Hohenester[30] predicted the strong-coupling regime's characteristic anti-crossing of energy levels for an emitter placed next to cigar-shaped nanoparticles.

In order to increase the Q-factor of the plasmonic modes so that the strong-coupling regime can be entered more easily, two main strategies have been pursued. The first concentrates on reducing the damping of the material. The high confinement and long lifetimes of graphene plasmons have been proposed in this regard[117]. In the second, cavities have been incorporated into plasmonic structures (Figures 3c and 3d). These plasmonic resonators combine the benefits of a high Q-factor and small mode volume[16,80,81,118,119,120]. De Leon *et al*. have proposed a plasmonic resonator composed of silver nanowires surrounded by dielectric Bragg reflectors[16] (Figure 3c), and demonstrated Purcell factors exceeding 75.

One of the main properties that make photons attractive for carrying quantum information is that they are weakly interacting. However, it also means that they do not interact with each other very well. Nonlinear materials can be used to boost this interaction, however the nonlinearity requires a high light intensity. This is unattractive as single-photon interactions are needed for quantum photonic devices. A strongly coupled light-emitter system has a nonlinear energy structure that allows for photon-photon interactions at the single-photon level (Box 3). In CQED this is known as the photon blockade[83]. An analogy has been found for plasmonics[121] and was used to devise the idea of a single-photon transistor[17,18,19]. As well as applications in photonics, the strong coupling regime in plasmonics has also been shown to be useful in the field of physical chemistry for enhancing chemical reactions[122].

In addition to single emitters, recent work has studied the interaction of multiple emitters mediated through a strong interaction with a plasmonic mode[123]. There have been predictions of a plasmonic Dicke effect where emitters coupled to a common plasmonic mode experience cooperative emission[124]. In a similar scenario, mediated interactions via a plasmonic mode generate entanglement between emitters[125,126,127]. This is a powerful insight as the proposed entanglement generation is induced from dissipative processes. In this way a perceived weakness of plasmonics has been converted into a positive.

**Nanolasers, metamaterials and many-body systems**

Despite the remarkable progress in studying light-matter interactions using plasmonic systems and a host of promising applications, the problem of high loss must be resolved for plasmonics to fulfil its full potential. Bergman and Stockman[128] have proposed a plasmonic version of a laser for providing amplification via stimulated emission. This 'spaser' could produce stimulated emission of SPPs by placing gain material around resonant metallic structures. The work paved the way for the creation and preservation of strong, coherent plasmonic fields at the nanoscale. Many proposals have since been put forward to exploit the spaser's novel effects, including the creation of subwavelength nanolasers, which out-couple the spaser's near-field as propagating radiation[129,130,131]. Due to the Purcell enhancement, these nanolasers can display threshold-less lasing[131]. Spasers have also been considered in the design of metamaterials to eliminate damping. As metamaterials have been brought from the microwave to the optical regime they have increasingly relied on plasmonic components[132]. Incorporating gain will be essential for the practical realization of their novel effects[6]. Metamaterials have also been considered for controlling quantum dynamics. Recent work has shown how negative-index metamaterials can aid non-linear interactions[133] and entanglement generation[134]. Experimental probing of metamaterials in the quantum regime has also been demonstrated[135].

One of the key successes of quantum optics over the last few decades has been the precise control of single quantum systems in a range of settings. Cold atom trapping in optical lattices, for instance, has helped shed light on a number of physical phenomena[136]. However, optical lattices are not easily scalable and the lattice period is restricted to half the wavelength of the trapping laser. Plasmonics has emerged as an alternative route towards investigating scalable solid-state systems for trapping atoms and molecules[137,138]. Due to the strong coupling between the emitter and the plasmonic mode, the metallic trap serves the dual purpose of trapping the atom as well as an efficient probe. The prospect of creating a plasmonic lattice with a nanometer period has been proposed[139]. These lattices would serve as an interesting playground to examine many-body physics in a parameter regime that is unavailable to traditional optical lattices.

## Quantum plasmonic circuitry

Plasmonic circuitry opens up a route toward nanophotonic quantum control with compact device footprints[1,2,37], enhanced coupling to emitter systems[17,18,19] and an electro-optical behaviour enabling interfacing with quantum photonic[2] and electronic components[140]. Quantum plasmonic circuitry can be decomposed into three principal stages[141]: (i) Generation, (ii) Manipulation, and (iii) Detection. The combination of these enables a self-contained 'dark' on-chip setting, where external far-field control is not required.

**Generation**

The generation of SPPs on waveguides has been achieved using various types of external quantum sources, including parametric down-conversion[43,45,46,47,48,50,51], an optical parametric oscillator[49] and emitters in cryostats[142,143]. A more integrated approach has been to embed emitters[11,12,16,94,95,96] directly on the waveguides and excite them with an external classical source, thereby generating single SPPs from the spontaneous emission. The high field confinement of the plasmonic mode enhances the process providing an efficient method to generate single SPPs. A more flexible approach, allowing the 'deterministic' launching of single SPPs was recently demonstrated by Cuche et al., where NV centres were fixed onto the tip-apex of a near-field optical microscope, enabling the generation of single SPPs at freely chosen positions[144]. By placing quantum dots in plasmonic cavities, enhanced SPP generation rates and frequency selectivity have been investigated theoretically[15] and observed experimentally[16]. The coupling of light from a single fluorophore molecule to a Yagi-Uda antenna structure has also been studied for generating single plasmonic excitations as coupled LSPs[13]. The generation of single LSPs at a silver nanostructure has recently been demonstrated[142]. In order to achieve truly integrated systems without external driving fields, however, the development of on-chip electrically driven SPP sources[145] will need to be pursued in the quantum regime.

**Manipulation**

In order to manipulate quantum states of SPPs, a range of waveguides have been considered for guiding, the most popular being nanowires[11,12,16,17,18,88,94,96]. While these provide a highly confined field that can be exploited for coupling the light to emitters, due to ohmic losses the propagation length - the distance the SPP field intensity drops to 1/e of its initial value - is small and of the order 10 $\mu$m at optical wavelengths. On the other hand, LRSPP waveguides have been probed in the quantum regime, where propagation lengths of up to 1 cm have been reported[45,46,49,51]. LRSPP waveguides provide relatively large propagation distances, but the field confinement is small. Thus, a combination of different waveguides may be required in order to reach an all-plasmonic solution for guiding. Materials such as graphene may also help reduce loss, while maintaining a high field confinement[38,117]. An alternative approach is a hybrid platform of metallic and dielectric waveguides, where the metal provides localized 'hotspots' for enhancing the coupling of light to emitters[10]. Another approach is using nanoparticles supporting coupled LSPs[146,147]. Several studies have used cavity-QED to investigate energy transfer[148], quantum state transfer[149], entanglement generation[150] and ultra-fast switching[19]. Finally, recent work has shown that by embedding gain material loss can be compensated in plasmonic waveguides in the classical regime[4]. Such techniques could potentially be used for guiding in the quantum regime.

More complex waveguide structures involving the quantum interference of SPPs have also been investigated. The excitation of SPPs into two directions on a gold nanowire provided the first realization of a quantum plasmonic beamsplitter[12]. The SPPs were also made to reflect at one end of the nanowire and propagate back to interfere in a Mach-Zehnder interferometer. More recent work has investigated the possibility of interfering two SPPs in continuous[151] and discrete[149] waveguides. Such a setting, if realized would show evidence of the bosonic nature of SPPs and initial experimental work has hinted at this[51]. Moreover, the interference of two SPPs forms an important first step in building up to more complex circuitry for the control of quantum states. In a more hybrid scenario, an integrated polarization sensitive beamsplitter has been designed[152].

**Detection**

The near-field detection of SPPs generated from a quantum dot source was demonstrated recently using a silver nanowire waveguide placed on top of a germanium field-effect transistor[96] (see Figure 4a). Here, the a.c. electric field of the SPP generates electron-hole pairs in the germanium nanowire. A d.c. electric field then separates these electron-hole pairs into free charges before recombination takes place. The separated electron-hole pairs are then detected as current, where a sensitivity of up to 50 electrons per SPP was detected. The detection of single SPPs propagating on gold stripe waveguides has also been demonstrated using superconducting nanowire detectors providing a faster operation and reduced dead-times[143] (see Figure 4b).

Future work on integrated generation, manipulation and detection, in addition to schemes for loss compensation promises to enable more complex nanoscale interference devices and coincidence-based quantum plasmonic operations to be realized.

# Perspectives

A huge amount of progress has been made in the growing field of quantum plasmonics. However, many quantum properties of surface plasmons are still to be fully explored and a number of problems remain along the route to realizing fully functioning and reliable quantum devices that take advantage of the intense light-matter interactions that plasmonics offers. The most pressing issue is how to deal with loss. While recent work has shown that loss compensation and gain can be achieved in basic plasmonic waveguides in the classical regime[4], it remains to be seen how these techniques can be translated into the quantum regime and in what way noise can be accommodated. It might be, however, that hybrid quantum plasmonic-photonic systems will be the optimal solution in the trade-off between confinement and loss[10,127,152], perhaps even exploiting the loss when needed for investigating dissipative effects in quantum systems[20]. Even with the problem of loss resolved, several important issues remain, such as fundamental limits on the quality of transistor-based quantum optical logic gates due to phase noise imparted on the signals during the nonlinear interaction[153]. Moreover, as we start to look at miniaturizing plasmonic components further, several questions are already beginning to appear: At what scale do current quantization methods based on a macroscopic approach break down? When will nonlocal microscopic effects, requiring density functional theory[56,62,64-77] combined with quantum optics, need to be considered in the design of new quantum plasmonic components? In Figure 5 we highlight some exciting and unexplored topics related to these questions. Finding the answers to these and many more related questions promises to make the next stage of research in the field of quantum plasmonics a very fruitful and productive time.

**Acknowledgements** We thank J. Takahara and C. Lee for comments on the manuscript. This work was supported by the UK's Engineering and Physical Sciences Research Council, the Leverhulme Trust, the European Office of Aerospace Research and Development (EOARD), the National Research Foundation of Korea grants funded by the Korean Government (Ministry of Education, Science and Technology; Grant numbers 2010-0018295 and 2010-0015059) and the Qatar National Research Fund (Grant NPRP 4-554-1-D84). SKO thanks L. Yang and F. Nori for their support.

**Competing interests statement** The authors declare that they have no competing financial interests.

**Corresponding Authors** Mark Tame (markstame@gmail.com) and Stefan Maier (s.maier@imperial.ac.uk).

**Figure 1. The Surface Plasmon Polariton (SPP).** The coupling of a photon and a plasmon at the interface of a material with a negative dielectric function (*e.g.* metal) and one with positive dielectric function (*e.g.* air) leads to a splitting of the ($\omega$-$k$) dispersion curves (solid lines) for the excitations which form a plasma shifted photon and a surface plasmon polariton as the joint state of light (photon) and matter (surface plasmon).

**Figure 2. Probing fundamental quantum properties of SPPs. a**, Plasmon-assisted transmission of polarization entangled photons through a metal grating consisting of a gold film perforated by an array of subwavelength holes[43]. The inset displays the fourth-order quantum interference fringes that show the entanglement survives the photon-SPP-photon conversion process. Here, the labels are BBO: beta-barium borate nonlinear crystal for photon generation via parametric down conversion, C: compensating crystal to adjust the phase between the components of the entangled state, HWP: half-wave plate, L: lens, TEL: confocal telescope, A1 and A2: metal grating, P1 and P2: polarizer, IF: interference filter, P1 and P2: single-photon detectors. **b**, Single SPPs excited in a metallic stripe waveguide by single photons from parametric down conversion[50] are found to preserve their photon-number statistics as witnessed by the second-order quantum coherence, $g^{(2)}(\tau)$ (inset). At the single-quanta level, SPPs are observed to experience loss consistent with an uncorrelated Markovian loss model, as suggested by the classical exponential behaviour of the count rates and the unchanged value of the second-order-coherence function with increasing waveguide length. **c**, Wave-particle duality of SPPs excited by a nitrogen vacancy (NV) centre in diamond placed in close proximity to a nanowire[12]. A single SPP interferes with itself (wave-like, top) and shows sub-poissonian statistics using a beamsplitter (particle-like, bottom). Here, the labels are NV: nitrogen-vacancy centre, BS: beamsplitter, d1 and d2 are the distance between the NV centre

and the close and far end of the nanowire, respectively, PA and PB are photodiodes, Pc is a photon correlator, A and B are the wire ends. **d**, Evolution of plasmonic modes as the inter-particle distance is varied from the classical regime through to the quantum regime[73]. The onset of quantum tunnelling determines a quantum limit of plasmonic confinement. Here, $d_{\mathrm{QR}}$ denotes the critical distance below which the plasmon interactions enter the quantum regime and R is the radius of the nanoparticle.

**Figure 3. Coupling of single emitters to SPPs.** In **a**, quantum dot emission into SPP modes of a silver nanowire is shown, as demonstrated in Akimov *et al's* experiment[11]. The dot can radiate into free space modes or SPP modes with rates $\Gamma_{rad}$ or $\Gamma_{pl}$, respectively. Alternatively, it can non-radiatively excite lossy surface modes, which quench the fluorescence. The quantum statistics of the fluorescence were investigated by observing the scattered light from the end of the nanowire. In **b**, the self-correlation coincidences of the scattered light from the SPP modes are shown. At $\tau = 0$, the coincidence counts almost reach zero. This indicates that the SPP mode scatters into single photons. The temporal width of the anti-bunching curve depends on the pumping rate R of a quantum dot from its ground state to excited state, and the total decay rate $\Gamma_{tot}$ back to the ground state. In **c**, a schematic (top) and scanning electron microscope image (bottom) of a plasmon distributed Bragg reflector resonator[16] is shown (scale bar 1 $\mu$m). In **d**, a sketch of a hybrid system of whispering gallery mode (WGM) in a microtoroid resonator and a metal nanoparticle (MNP) cavity is shown, including a zoomed in view of the nanoparticle and the emitter[118]. Here, $\kappa_0$ is the intrinsic damping of the WGM, $\kappa_1$ is the coupling between the tapered fibre and the WGM, $r_m$ is the radius of the MNP, $d$ is the distance between the MNP and the emitter, $G$ is the vacuum Rabi frequency of the emitter, $\gamma_s$ is the spontaneous emission rate of the emitter, $\kappa_R$ is the radiative damping rate of the WGM due to scattering from the MNP and $K_m$ is the ohmic damping rate of the MNP. In this composite resonator the nanoparticle acts as an antenna efficiently coupling the emitter into the high-Q WGM cavity.

**Figure 4. Quantum plasmonic circuitry.** In **a** and **b**, two approaches are shown that have recently been used for realizing on-chip detection of single SPPs. In **a**, electron-hole pair production in a germanium nanowire (inset), from the light field of SPPs on the silver nanowire, generates a current that can be used for detection[96]. In **b**, superconducting nanowire detectors are placed on top of gold stripe waveguides in order to achieve single-SPP detection in the near field[143]. Here, the waveguide splits to form an integrated plasmonic Hanbury-Brown and Twiss interferometer that is used to demonstrate antibunching of single SPPs by measurement of the second-order coherence. In **c**, a hybrid metal-dielectric beamsplitter is shown, where the excitation of SPPs enables an integrated polarization sensitive beamsplitter[152] that can be used for quantum information processing. Here unique properties of SPPs are used, including high-field confinement (compactness), polarization dependence (variable splitting) and broadband nature (fast operation).

**Figure 5. Quantum plasmonics roadmap.** A range of topics on the horizon for the field of quantum plasmonics. This includes the development of new quantum plasmonic applications, such as 'dissipative driven quantum dynamics' and probing deeper into the fundamental properties of light-matter systems, such as their microscopic quantization and their potential for ultra-strong quantum interactions with emitters at the nanoscale.

**Box 1. Quantization of surface plasma waves.** A basic approach to the quantization of surface plasma waves (SPWs) involves quantizing the electromagnetic field by accounting for the dispersive properties of the metal via the collective response of the electrons[32,33]. A mathematically equivalent, but more rigorous approach including loss uses the microscopic Hopfield[27] or macroscopic Green's function[28] formalism. There are 3 steps to quantization: (i) Classical mode description, (ii) Discretization of classical modes, and (iii) Quantization via the correspondence principle. We briefly present these steps for SPPs and LSPs.

**(i) Classical mode description:** In Figure B1a the SPW at a plane interface between a metal and vacuum (or air) is shown. The metal has a dielectric constant $\varepsilon(\omega)$ and initially loss is neglected. One can describe the total electromagnetic field in terms of a vector potential, $\boldsymbol{A}(\boldsymbol{r},t)$, where the electric and magnetic fields are recovered in the usual way using Coulomb's gauge ($\nabla \cdot \boldsymbol{A} = 0$), i.e. $\boldsymbol{E} = -\frac{\partial \boldsymbol{A}}{\partial t}$ and $\boldsymbol{B} = \nabla \times \boldsymbol{A}$. By solving Maxwell's equations a general form of the vector potential for the SPW is found to be

$$\boldsymbol{A}(\boldsymbol{r},t) = \frac{1}{(2\pi)^2} \int d^2K \alpha_K \boldsymbol{u}_K(\boldsymbol{r}) \exp(-i\omega t) + \text{c.c.}$$

Here, c.c denotes the complex conjugate, $\boldsymbol{K}$ is a real wave vector parallel to the interface and the frequency $\omega$ is linked to the wavenumber, $K = |\boldsymbol{K}|$, by the dispersion relation $K = \frac{\omega}{c}\sqrt{\frac{\varepsilon(\omega)}{\varepsilon(\omega)+1}}$. In addition, the term $\alpha_K$ is an amplitude and the mode function $\boldsymbol{u}_K(\boldsymbol{r})$ is given by

$$\boldsymbol{u}_K(\boldsymbol{r}) = \frac{1}{\sqrt{L(\omega)}} \exp(-\kappa_j z)(\widehat{\boldsymbol{K}} - i\frac{K}{\kappa_j}\widehat{\boldsymbol{z}})\exp(i\boldsymbol{K} \cdot \boldsymbol{r}),$$

where $L(\omega)$ is a length normalization and $\kappa_j^2 = K^2 - \varepsilon_j \omega^2/c^2$ characterizes the decay of the field in the $z$ direction ($\pm ve$ solution of $\kappa_{1,2}$ chosen for $\pm z$, respectively), with $\varepsilon_1 = 1$ and $\varepsilon_2 = \varepsilon(\omega)$. In the above, $\widehat{\boldsymbol{v}}$ denotes the unit vector for the vector $\boldsymbol{v}$.

**(ii) Discretization of classical modes:** To discretize the SPW mode functions, a virtual square of area $S = L_x \times L_y$ is introduced on the surface. This gives discretized values for the wavenumbers $K_x = n_x 2\pi/L_x$ and $K_y = n_y 2\pi/L_y$, where $n_x$ and $n_y$ are integers. By substituting $\frac{1}{(2\pi)^2} \int d^2K \to \frac{1}{S}\Sigma_K$ and $\alpha_K \to SA_K$ one obtains a discretized form for $A(r,t)$. Using the formula for the total energy of the electromagnetic field in the virtual square $U = \int dt \int dr (E \frac{\partial D}{\partial t} + H \frac{\partial B}{\partial t})$, where $D = \varepsilon_0 E + P = \varepsilon_j E$ and $H = \mu_0 B$ are used, one finds

$$U = \sum_K \varepsilon_0 \omega^2 S [A_K A_K^* + A_K^* A_K]$$

which has exactly the structure of the energy of a harmonic oscillator for each mode $K$.

**(iii) Quantization via the correspondence principle:** Using the quantized Hamiltonian of a harmonic oscillator $\hat{H} = \sum_K \frac{\hbar \omega_K}{2}[\hat{a}_K \hat{a}_K^\dagger + \hat{a}_K^\dagger \hat{a}_K]$ with the correspondence $A_K \to \sqrt{\frac{\hbar}{2\varepsilon_0 \omega S}} \hat{a}_K$ and $A_K^* \to \sqrt{\frac{\hbar}{2\varepsilon_0 \omega S}} \hat{a}_K^\dagger$, the field of the SPW is quantized by the association of a quantum mechanical oscillator with each mode $K$. The operators $\hat{a}_K$ and $\hat{a}_K^\dagger$ are annihilation and creation operators which destroy and create a quantum of energy, $\hbar\omega_K$, and obey bosonic commutation relations $[\hat{a}_K, \hat{a}_{K'}^\dagger] = \delta_{K,K'}$. A single quantized SPW excitation, or SPP (now both a wave and a particle), is then written as $|1_K\rangle = \hat{a}_K^\dagger |\text{vac}\rangle$, where $|\text{vac}\rangle$ represents the vacuum state of the system. The commutation relations are responsible for the different behaviour of physical observables compared to the classical regime. For example, the widely used second-order coherence function at a fixed position[34], $g^{(2)}(\tau) = \frac{\langle \hat{E}^-(0)\hat{E}^-(\tau)\hat{E}^+(\tau)\hat{E}^+(0)\rangle}{\langle \hat{E}^-(0)\hat{E}^+(0)\rangle^2}$, quantifies the probability of measuring an excitation at time $t = 0$ and another at $t = \tau$. Here, $\langle \hat{X} \rangle$ represents the expectation value of the operator $\hat{X}$ and $\hat{E}^+(t)$ ($\hat{E}^-(t)$) is the positive (negative) electric field - a function of annihilation (creation) operators. For single SPPs, $g^{(2)}(0) = 0$, whereas for SPWs, as there is no commutation the numerator factorizes to give $g^{(2)}(0) \geq 1$. Examples of $g^{(2)}(0)$ can be seen in Figures 2b and 3b.

The above quantization procedure can be carried out for more complex waveguides, such as channel and nanowires, with the only change being the mode function $\boldsymbol{u}_K(\boldsymbol{r})$ which represents the classical wavelike properties of the excitation. In most cases a continuum limit is used for the wavevector $\boldsymbol{K}$. In order to include loss in the quantization, one couples the SPP to a reservoir of bath modes[33], $\hat{b}_i$, as depicted in Figure B1a, whose coupling strength is determined in a phenomenological approach from the imaginary part of $\varepsilon(\omega)$ for the metal, which is a result of the damping experienced by the electrons. This is mathematically equivalent to the more rigorous reservoir method[27].

A similar procedure is used to quantize the near field of localized plasma oscillations at nanoparticles[31], as shown in Figure B1b. The vector potential for the field can be written as $\boldsymbol{A}(\boldsymbol{r},t) = \sum_i \alpha_i \boldsymbol{u}_i(\boldsymbol{r}) \sin(\omega_0 t)$, where the mode function is given by $\boldsymbol{u}_i(\boldsymbol{r}) = \hat{\boldsymbol{i}}$ for $r < R$ and $\boldsymbol{u}_i(\boldsymbol{r}) = -\frac{R^3}{r^3}[3(\hat{\boldsymbol{i}} \cdot \hat{\boldsymbol{r}})\hat{\boldsymbol{r}} - \hat{\boldsymbol{i}}]$ for $r > R$. Here, the subscript $i$ represents the three-dimensional coordinates ($i = x, y, z$), $R$ is the radius of the nanoparticle and $r$ is the radial coordinate of the position vector $\boldsymbol{r}$, taken with respect to the centre of the nanoparticle. Following similar steps as for SPWs, one obtains bosonic annihilation and creation operators $\hat{a}$ and $\hat{a}^\dagger$, where $|1\rangle = \hat{a}^\dagger|\text{vac}\rangle$ represents a single quantized localized surface plasma oscillation, or localized surface plasmon (LSP), corresponding to the creation of a quantum of energy $\hbar\omega$ in the near field of the nanoparticle. Internal damping is then modeled as a reservoir of bath modes, $\hat{b}_i$, as in the SPP case. The far-field radiation can also be treated as a reservoir of bath modes, $\hat{c}_i$, the evolution of which can be tracked and measured if desired, such that they do not constitute a fundamental loss channel. A more rigorous approach[30,35] enables the treatment of LSPs at arbitrary shaped nanostructures and with hydrodynamic effects included.

**Figure B1. Quantization of SPPs on waveguides and LSPs at nanoparticles.** In **a**, the magnitude of the transverse $z$ component of the mode function $\boldsymbol{u}_K(\boldsymbol{r})$ for a single SPP excitation, denoted by $\hat{a}$, is shown along with a selection of alternative waveguide geometries. In **b**, the magnitude of the radial $r$ component of the mode function $\boldsymbol{u}_i(\boldsymbol{r})$ for a single LSP excitation is shown. Reservoir modes, denoted by $\hat{b}$ and $\hat{c}$, are also shown for both SPPs and LSPs, enabling loss to be included in the quantization.

**Box 2. SPP field confinement: subwavelength vs subdiffraction.** The field associated with SPP quanta can be highly confined to both subwavelength and subdiffraction dimensions. In order to see this, one can consider the $k$-space surfaces for three different scenarios, as shown in Figure B2.

For light in a bulk 3-dimensional (3D) material with positive $\varepsilon$, as shown in Figure B2a (bottom), the spatial spread of a beam in a plane ($y$-$z$) transverse to the direction of propagation ($x$) must satisfy $\Delta k_i \Delta i \geq 2\pi$, where $i = y, z$, $\Delta i$ is the spatial spread in direction $i$ and $\Delta k_i$ the corresponding spread in wavenumber. This inequality is due to the Fourier reciprocity that occurs when arbitrary fields are expanded as a synthesis of plane waves. Using Maxwell's equations one finds the relation between the wavenumber components, $k_x^2 + k_y^2 + k_z^2 = k_0^2 \varepsilon = k^2$, where $k_0$ is the freespace wavenumber, $k_0 = \frac{2\pi}{\lambda_0}$, and $\lambda_0$ is the freespace wavelength. The $k$-space surface for 3D waves is shown in Figure B2a (top), where the maximum variation of a wavenumber is $\Delta k_i \leq 2k$. This leads to the well-known 3D diffraction limit, $\Delta i \geq \frac{\lambda_0}{2n}$. Subwavelength confinement (compared to $\lambda_0$) can be achieved by simply using a material with a larger refractive index, $n = \varepsilon^{1/2}$.

At an interface between two materials with positive dielectric functions, $\varepsilon > \varepsilon_2$, as shown in Figure B2b (bottom), where total internal reflection takes place, *e.g.* at the interface between core and cladding in an optical fibre or a cavity wall, one finds $k_x^2 + k_y^2 + k_z^2 = k_0^2 \varepsilon$ and $k_x^2 + k_y^2 - \kappa_z^2 = k_0^2 \varepsilon_2$. Here, the $z$ component of the wavevector in the upper material has become imaginary ($k_z = i\kappa_z$, with $\kappa_z$ real) representing the evanescent decay of the field. The $k$-space surfaces for these two equations are shown in Figure B2b (top). As $k_x^2 + k_y^2$ must match across the interface one finds the 3D diffraction limit still applies in the $x$-$y$ plane in the upper material, with an evanescent decay given by an exponential function with $1/e$ length $\delta_D = 1/\kappa_z \gtrsim \frac{\lambda_0}{2n}$, which can be subwavelength for large $n$.

By replacing the lower material with one that has a negative dielectric function, as shown in Figure B2c (bottom), one is able to 'break' the 3D diffraction limit. Here, noble metals such as gold can be used, where the effective response of the electrons at the surface to the coupled field can be described by a Drude-Lorentz dielectric function[26], $\varepsilon(\omega)$, which is negative for frequencies below the plasma frequency. In this negative regime, using Maxwell's equations, one finds $k_x^2 + k_y^2 - \kappa_z^2 = k_0^2 \varepsilon_2$ and $k_x^2 + k_y^2 - \kappa_z'^2 = k_0^2 \varepsilon(\omega)$. Here, the $z$ components of both wavevectors have become imaginary – the field has become 2D. The $k$-space surfaces of these two equations that represent the combined light field supported by the electrons (the SPP) are shown in Figure B2c (top). While $k_x^2 + k_y^2$ must match across the boundary, its value is no longer limited, which in principle enables confinement to arbitrary spatial extent in the $x$-$y$ plane. However, an additional constraint comes from the maximum value that $k_x^2 + k_y^2$ can take, given by the dispersion relation for the SPP, as shown in Figure 1. For the geometry considered we have $k_{SPP} = \sqrt{k_x^2 + k_y^2} = \frac{\omega}{c}\sqrt{\frac{\varepsilon_2 \varepsilon(\omega)}{\varepsilon_2 + \varepsilon(\omega)}}$. From Fourier reciprocity this gives $\Delta x, \Delta y \geq \frac{\lambda_0}{2n}\sqrt{1 - \frac{\varepsilon_2}{|\varepsilon(\omega)|}}$ and $\delta_D = 1/\kappa_z \geq \frac{\lambda_0}{2n} f(\varepsilon_2, \varepsilon(\omega))$. Both can be made significantly smaller than their positive dielectric counterparts. The amount depends on the materials and geometry, with nanowires and channel waveguides providing even larger field confinement[36,37].

**Figure B2. $k$-space surfaces.** In **a**, the $k$-space surface for a photon in a bulk 3D material is shown. One can see the maximum spread for any wavenumber is $2k$, leading to the diffraction limit. In **b**, the $k$-space surfaces for a 2D photon are shown, where total internal reflection has taken place. Here, the total transverse wavenumber, $k_x^2 + k_y^2$, must match across the interface so that the maximum spread of the individual wavenumbers $k_x$ and $k_y$ in the upper material is again diffraction limited. In **c**, the $k$-space surfaces for the field associated with an SPP are shown, where the total transverse wavenumber is no longer diffraction limited. Its value now depends on the waveguide geometry and material used.

**Box 3. Weak and strong coupling in plasmonic cavity quantum electrodynamics.** The spontaneous emission of an emitter is strongly dependent on the electromagnetic environment it resides in[41]. Cavity quantum electrodynamics (CQED) studies the interaction of emitters with tailored electromagnetic fields[34]. Typically these fields are monomodes with high quality factors (Q) and small effective volumes ($V_{eff}$). These properties provide emission enhancement, which is formally defined by the Purcell factor

$$F_p = \frac{\gamma_{cavity}}{\gamma_{free\,space}} \propto Q\left(\frac{\lambda^3}{V_{eff}}\right),$$

where $\gamma$ is the decay rate of the emitter. The strength of the interaction between the emitter and the field is characterized by a coupling frequency, $g \propto \sqrt{\frac{1}{V_{eff}}}$. CQED can be split into two regimes which are dependent on the comparison of $g$ and the damping rates of both the emitter and the cavity ($\gamma, \kappa$). These regimes are classified as the weak-coupling regime ($g \ll \gamma, \kappa$) and the strong-coupling regime ($g \gg \gamma, \kappa$) [74].

CQED has been a popular platform for proof-of-principle implementations of quantum information processing[78]. However, the diffraction-limited optical cavities place a lower bound on the size of these systems. The drive to bring CQED down to the nanoscale has opened the door to plasmonic CQED. Here, both SPP and LSP plasmonic modes offer subwavelength and subdiffraction field confinement that enables extreme light-matter coupling. In particular, the resonant LSP modes supported by metal nanoparticles can be described effectively as a leaky cavity in quantum optics formalism, as shown in Figure B3a. Recent work on adding resonators to waveguide SPP systems brings these types of modes into the quasimode regime of CQED as well[16,79,80,81].

In the weak-coupling regime the excitation within the emitter-cavity system is irreversibly lost to the outside environment before any coherent exchange of energy can occur. In this case the light field has a perturbative effect on the emitter, which manifests itself as a modification of the decay rate as described above. In the strong-coupling regime a reversible exchange of energy, known as Rabi oscillations, exist between the emitter and the cavity field[82], as shown in Figure B3b for an emitter coupled to a metal nanoparticle. At this point the subsystems can no longer be treated separately. The composite system is described as a dressed emitter whose eigenenergies show a degeneracy splitting in comparison to the undressed emitter[34], as shown in Figure B3c. The dressed emitter's energy spacings provide a nonlinearity that enables single-photon nonlinear optics via the photon blockade phenomena[83].

**Figure B3. Weak and strong coupling.** In **a**, the analogy is shown between an atom in a single mode leaky cavity (left) and an atom residing in the near-field of a resonant LSP mode supported by a metal nanoparticle (right). The principal difference between the two is their dissipation channels. The cavity loses photons by transmission through its side walls. The LSP mode on the other hand is dissipated through radiative losses and ohmic losses associated with the metal nanoparticle. In **b**, the Rabi oscillation describes the transfer of an excitation between the LSP mode and the emitter at a Rabi frequency, $g$. In **c**, a schematic of the dressed emitter's energy levels is shown. Each energy manifold of excitation number $N$ has two states associated with it, $|N, +\rangle$ and $|N, -\rangle$. The magnitude of the difference between the dressed emitter and undressed emitter energy levels for equal $N$ is given by $g\sqrt{N}$. This anharmonic splitting is shown in the diagram and explains how a photon of a certain frequency may only excite the $N = 1$ manifold and nothing greater. *i.e.* a photon blockade.

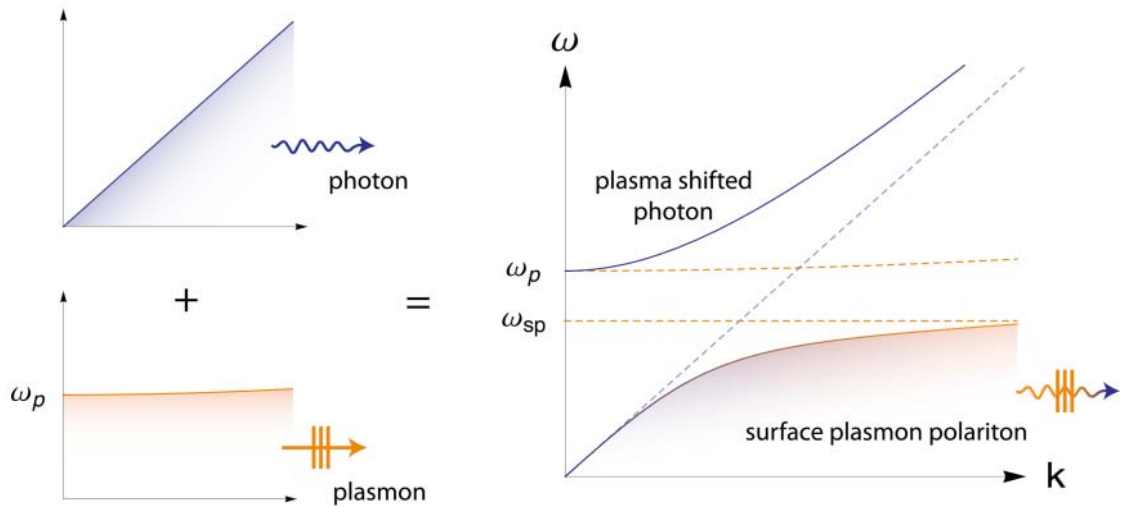

**Figure 1**

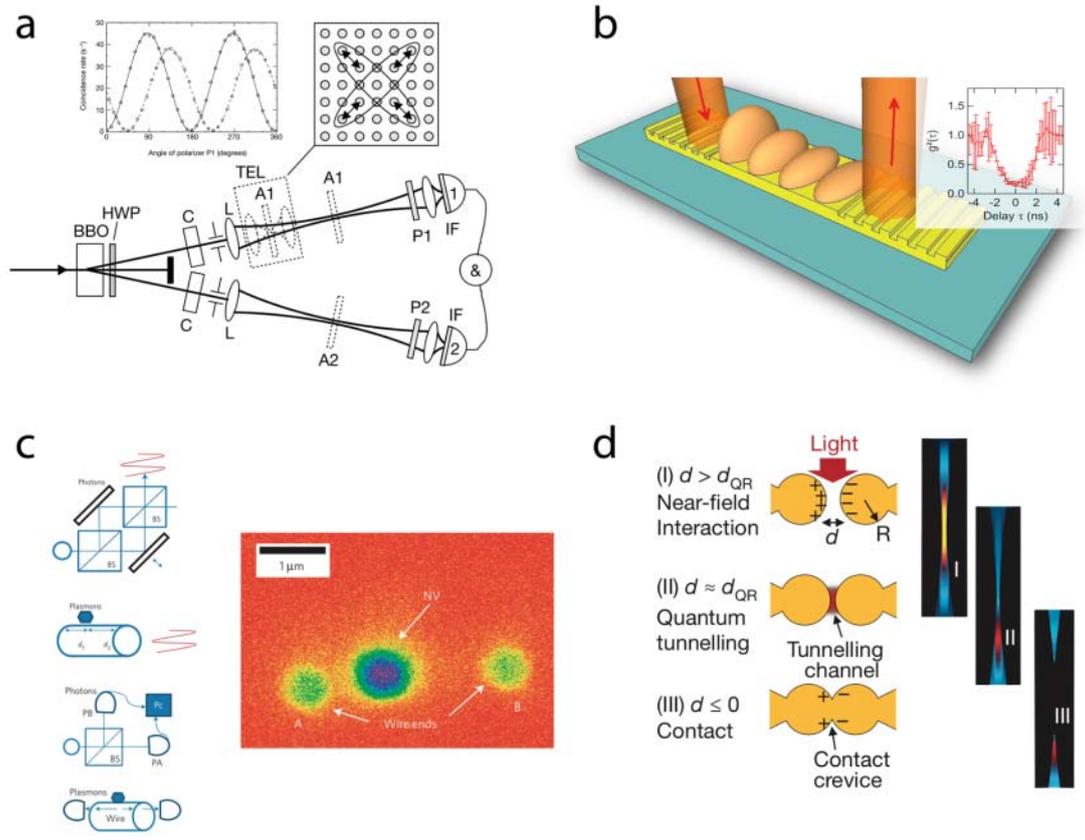

**Figure 2**

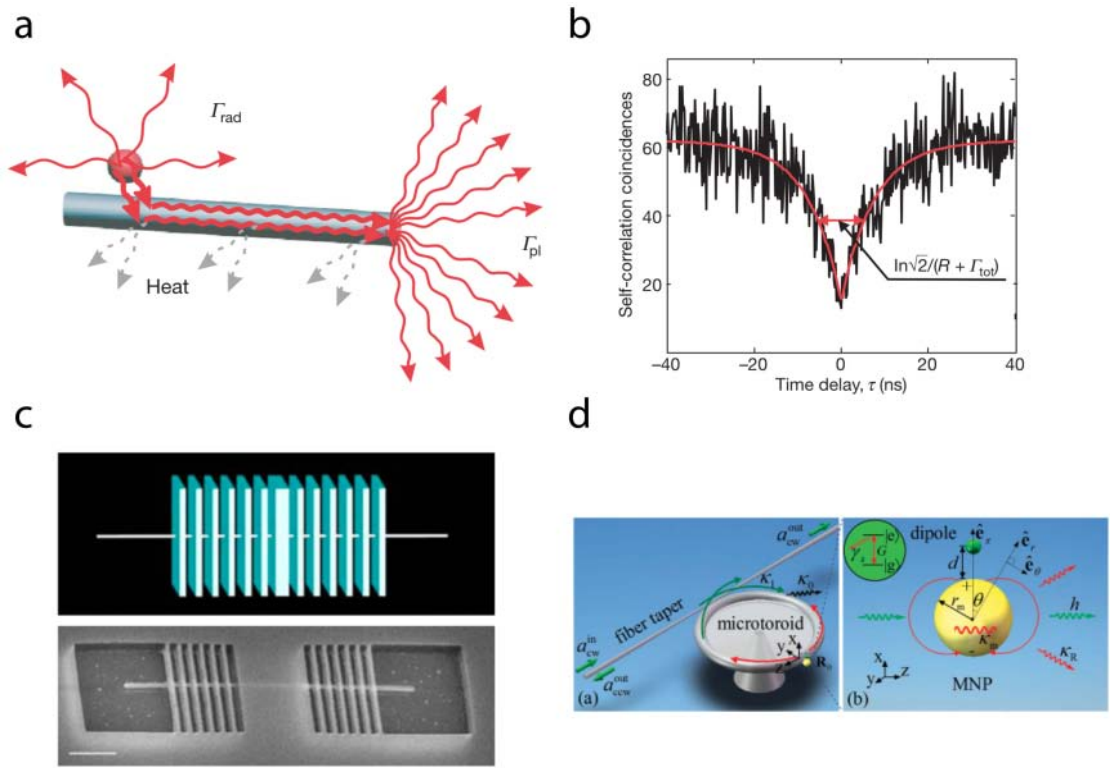

**Figure 3**

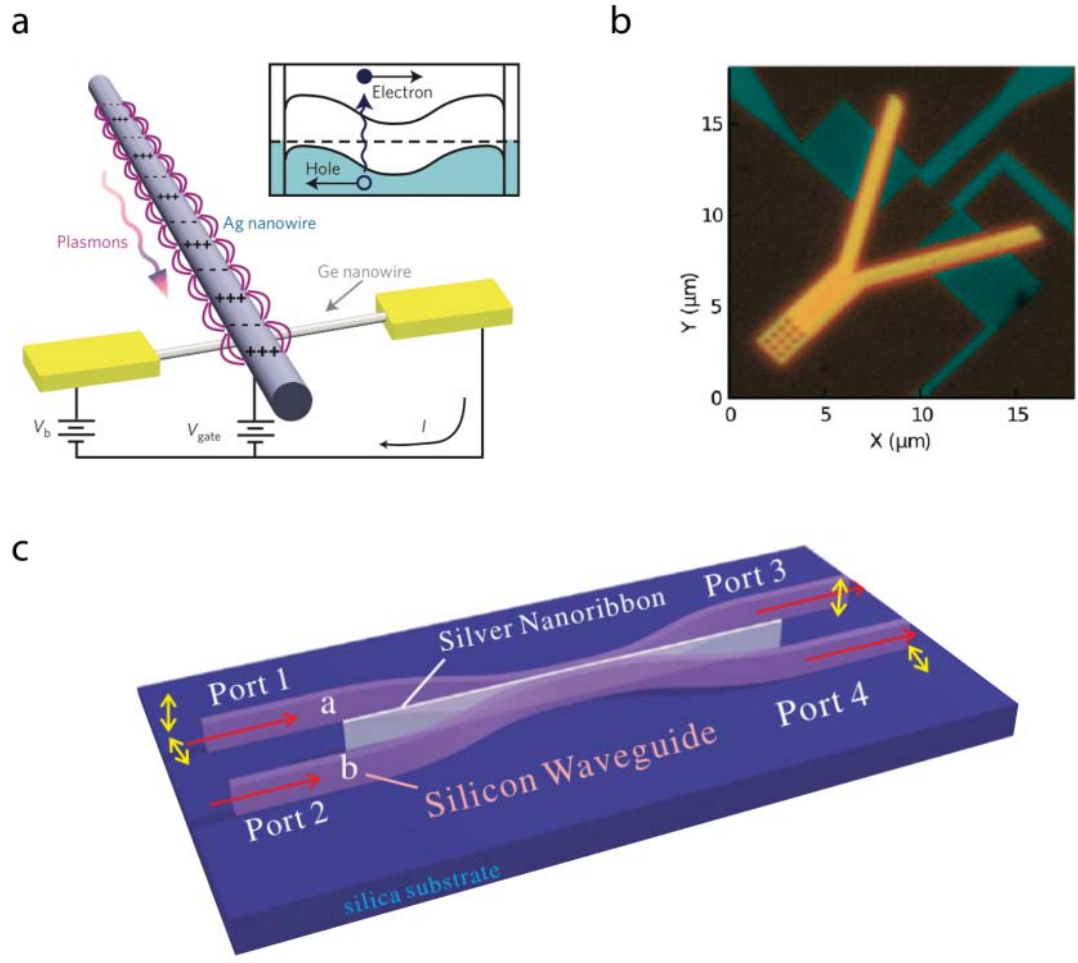

**Figure 4**

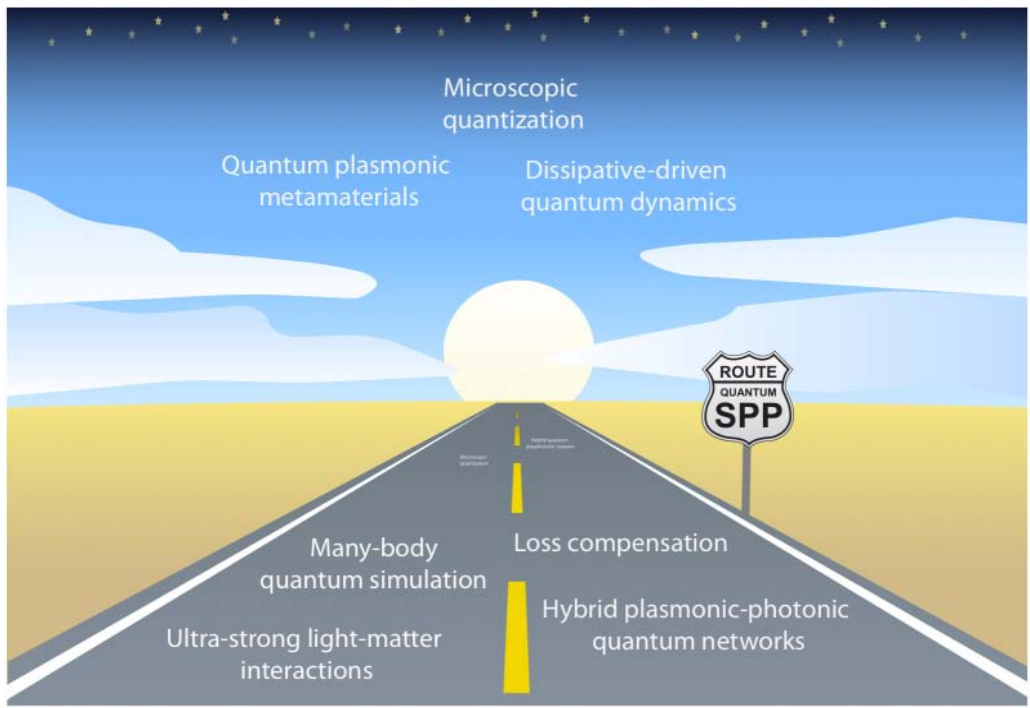

**Figure 5**

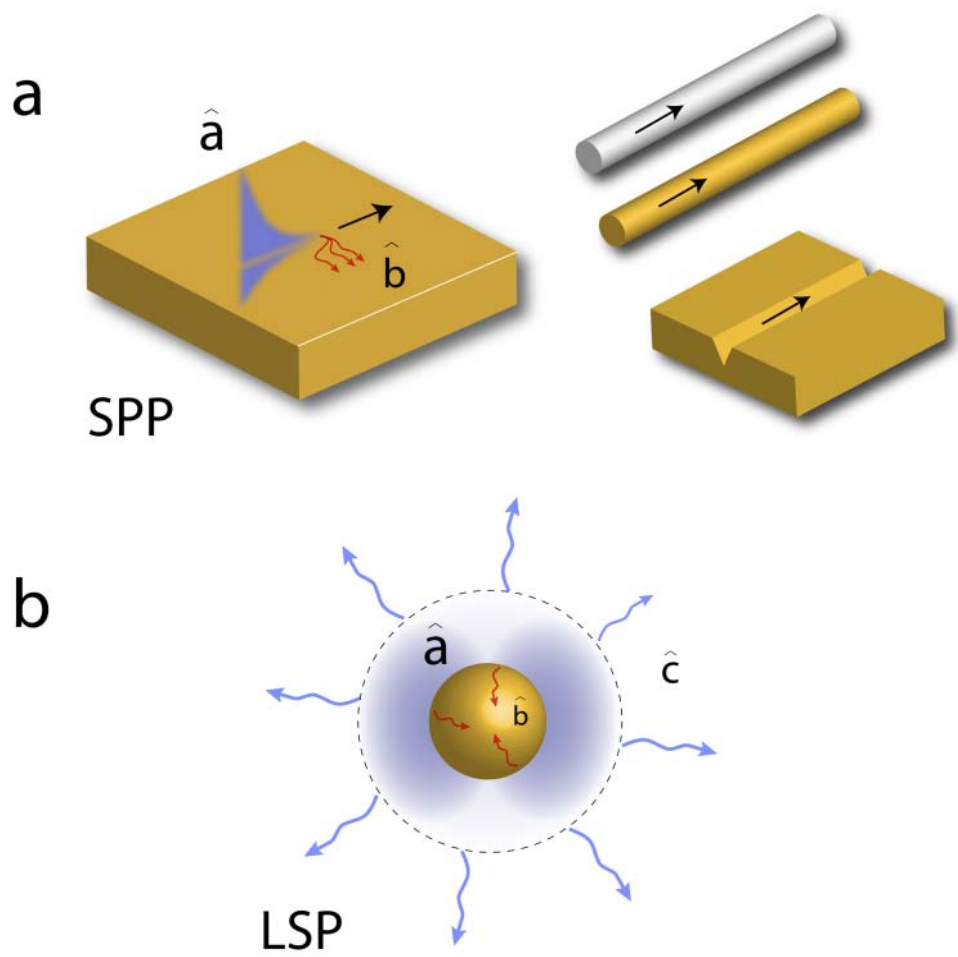

**Figure B1**

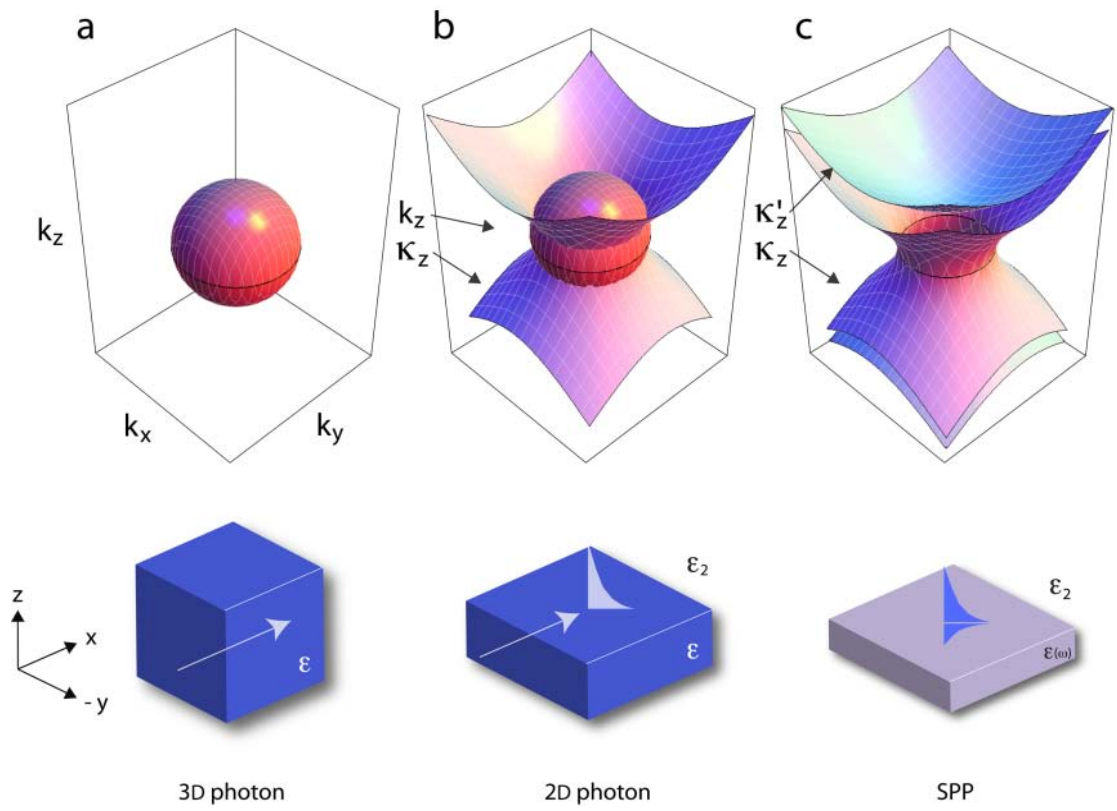

**Figure B2**

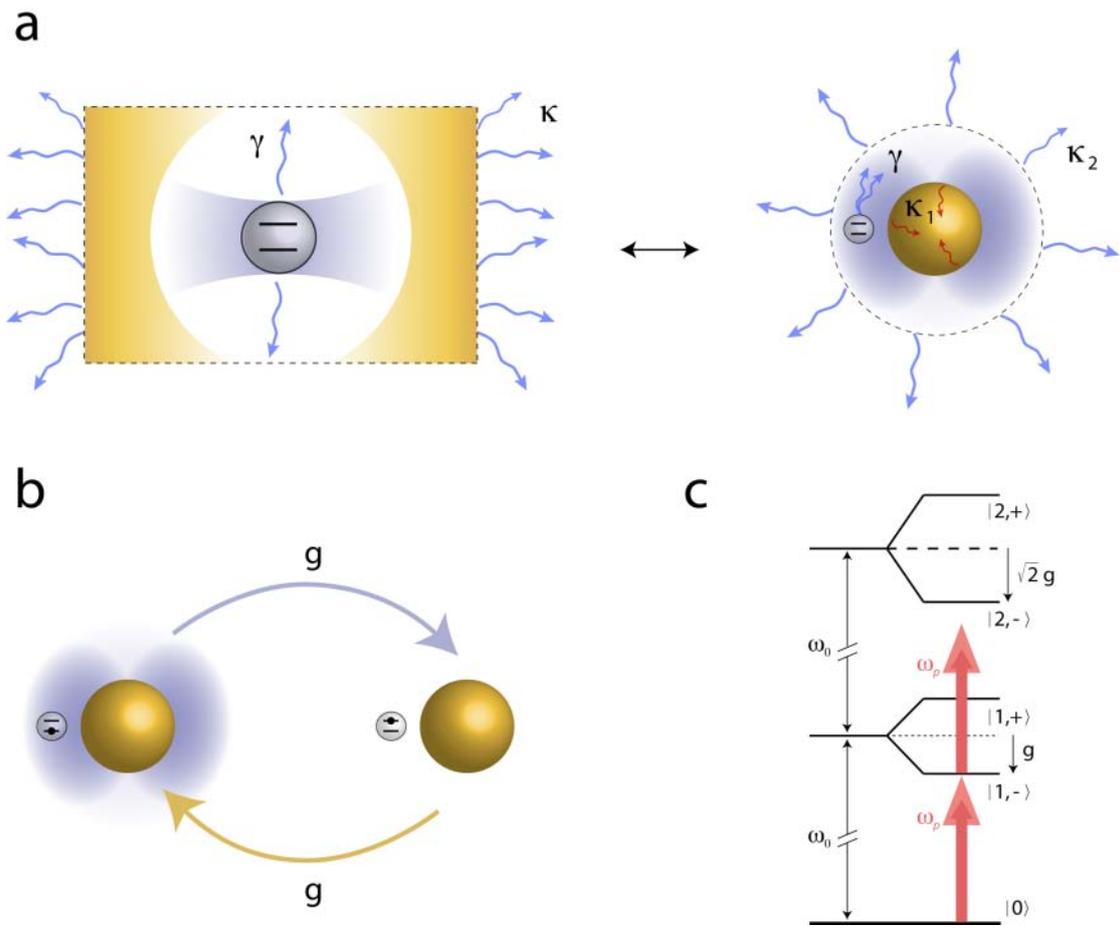

**Figure B3**